\documentclass[12pt,preprint]{aastex}
\usepackage{graphicx}
\usepackage[final]{epsfig}

\shorttitle{Relativistic Dyson Rings}
\shortauthors{Ansorg, Kleinw\"{a}chter and Meinel}

\begin{document}

\title{Relativistic Dyson Rings and Their Black Hole Limit}

\author{M. Ansorg\altaffilmark{1}, A. Kleinw\"{a}chter\altaffilmark{2}, 
        and R. Meinel\altaffilmark{3}}

\affil{Theoretisch-Physikalisches Institut,\\ University of Jena,
       Max-Wien-Platz 1, 07743 Jena, Germany}
\altaffiltext{1}{Electronic mail: Ansorg@tpi.uni-jena.de}
\altaffiltext{2}{Electronic mail: Kleinwaechter@tpi.uni-jena.de}
\altaffiltext{3}{Electronic mail: Meinel@tpi.uni-jena.de}


\begin{abstract}
In this Letter we investigate uniformly rotating, 
homogeneous and axisymmetric relativistic fluid bodies
with a toroidal shape. The corresponding field equations
are solved by means of a multi-domain spectral method, which
yields highly accurate numerical solutions. For a prescribed, sufficiently large ratio
of inner to outer coordinate
radius, the toroids exhibit a continuous transition to the extreme Kerr black
hole. Otherwise, the most relativistic configuration rotates at the
mass-shedding limit.
For a given mass-density, there seems to be no bound to the gravitational mass
as one approaches the black-hole limit and
a radius ratio of unity.
\end{abstract}



\section{Introduction}

Self-gravitating toroidal fluid configurations (without a central body)
in Newtonian Gravity were the subject of analytic investigations 
by \citet{Poi85a,Poi85b,Poi85c,Dys92,Dys93,Kow95} and \citet{Lic33}. In particular, Dyson
was able to give a fourth order expansion of uniformly rotating, 
homogeneous and axisymmetric rings, which turned out to be an extremely
good approximation for thin rings, see \citet{An02b}.
Numerical evidence for the existence of these rings was given by
\citet{Won74} and later by \citet{Eri81}, who, in addition, confirmed a
conjecture by \citet{Bar71} stating that there is a continuous
connection between the Maclaurin spheroids and the sequence of `Dyson
rings', see also \citet{An02b}.

In this Letter we extend the above Dyson rings to
Einsteinian Gravity. As one moves away from the Newtonian
configurations, one observes typical relativistic effects such as
the formation of ergo-regions. In particular, we found an interesting
continuous transition to an extreme Kerr black hole. If a fixed ratio
$\rho_1/\rho_2>0.5613$ of inner to outer coordinate radius is
prescribed and one gradually increases the gravitational mass (for
fixed mass-density), the
configurations eventually form an extreme Kerr black hole, as
described by \citet{Mei02}. If on the other hand
$\rho_1/\rho_2<0.5613$ is fixed, the ultimate configuration, as one
increases the gravitational mass, rotates at the mass-shedding limit.
It is furthermore interesting to note that the maximum gravitational mass
seems to become infinite as $\rho_1/\rho_2\to 1$.

For computing the `relativistic Dyson rings' we extended our
multi-domain spectral method \citep{An02a} to toroidal topology.
Again we obtain an accuracy of up to 12 digits for configurations
sufficiently far away from limiting cases.

Relativistic Dyson rings and their possible generalizations (rings with realistic
equations of state, configurations with a central object) might be relevant in two
different astrophysical situations: (i) they could form as a result of stellar
core-collapse in the case of high angular momentum, cf. \citet{Tho98}; (ii) they
could be present in central regions of galaxies. Of course, for a final evaluation
of the astronomical relevance of relativistic toroids their stability must be analyzed.
It is to be expected that they are stable with respect to axisymmetric perturbations,
but they may be unstable to (non-axisymmetric) fragmentation. In this case, the toroids
could nevertheless play an interesting role as intermediate configurations in various
astrophysical collapse scenarios.

In what follows, units are used in which the speed of light as
well as Newton's constant of gravitation are equal to 1.

\section{Metric Tensor, Field Equations and Boundary Conditions}

For an axisymmetric and stationary space-time describing the gravitational field of
a uniformly rotating perfect fluid body the line element
can be cast into the following form\footnote{This form of the line element of an
axisymmetric and stationary space-time also applies in more general situations,
e.g. for differentially rotating perfect fluids or if the stress-energy tensor is that of
a stationary, axisymmetric electromagnetic field \citep{Ca69}.}:
\[ds^2={\rm e}^{2\alpha}(d\rho^2+d\zeta^2)
       +W^2{\rm e}^{-2\nu}(d\varphi-\omega\, dt)^2
        -{\rm e}^{2\nu}dt^2\,.\]
We define the corresponding  
Lewis-Papapetrou coordinates $(\rho,\zeta,\varphi,t)$ uniquely
by the requirement that the metric coefficients and their first derivatives 
be continuous at the surface of the body. 

A particular consequence of the interior field equations for a perfect fluid body
revolving with the uniform angular velocity $\Omega$ is the
boundary condition
\[{\rm e}^{2\nu}-W^2(\omega-\Omega)^2{\rm e}^{-2\nu}={\rm
const.}=(1+Z_0)^{-2}\, , \]
which holds along the surface of the fluid. The constant is
related as shown to the relative redshift $Z_0$, measured at infinity, of photons
which are emitted from the body's surface and do not carry angular
momentum. 

Interior and exterior field equations together with
the above boundary and transition conditions at the body's surface, 
asymptotic behaviour at infinity
and regularity conditions along the rotation axis ($\rho=0$) form a complete
set of equations to be solved, see for example \citet{BI76}.

As long as we do not consider the transition to the extreme Kerr black
hole, the regularity condition at infinity is asymptotic flatness. 
If we however follow the `physical route' to the extreme Kerr black hole
studied by \citet{Mei02}, we learn that in this limiting process the fluid body
shrinks until it coincides with the coordinate origin. The geometry
of the exterior space-time assumes that of the extreme Kerr solution outside the horizon.
On the other hand, a completely different space-time, which is not asymptotically
flat, forms if we rescale the coordinates such that the fluid body retains
its finite extension. The asymptotic behaviour of the corresponding 
gravitational potentials in this limit is given by the
`extreme Kerr throat geometry', see \citet{BH99} and \citet{Mei02}. In order to
determine the physical parameters of the rings in the black-hole limit precisely,
it is necessary to calculate this inner solution
with its non-flat asymptotic behaviour.

\section{The Multi-Domain Spectral Method}

In the case of spheroidal figures of equilibrium, a two-domain
spectral method was used to yield highly accurate numerical
solutions \citep{An02a}. In this method we separately mapped the interior and
exterior of the star onto a square. The field quantities as well 
as the unknown shape of the fluid body were written in terms of
Chebyshev expansions, and the corresponding coefficients
resulted from a high-dimensional nonlinear set of equations
that incorporated both field equations and transition
conditions. This system was solved by a Newton-Raphson method,
and an initial guess for the solution was taken from the
analytically known Newtonian Maclaurin spheroids.

The idea of mapping several subregions separately onto squares
can also be applied to the case of toroidal figures. In a first step
we map the interior of the ring onto a square
$(0\leq s\leq 1, 0\leq t\leq 1)$ by
\[ \begin{array}{lll}
        \rho^2 &=&\rho_1^2+(\rho_2^2-\rho_1^2)s\\[2mm]
        \zeta^2&=&(1-t)y_{\rm B}(s)
   \end{array} \]
where $\rho_{1/2}$ stands for the inner and
outer coordinate radius of the ring, respectively,
and the non-negative function $y_{\rm B}$, which describes the unknown shape of the
ring's surface, satisfies
\[y_{\rm B}(0)=0\,,\quad y_{\rm B}(1)=0. \]
(We assume reflectional symmetry
with respect to the plane $\zeta=0$.)
Next we
introduce toroidal coordinates $(\tilde{\rho},\tilde{\zeta})$
in order to obtain a compact
coordinate region for the entire space exterior to the ring:
\[  z=\mbox{i}\rho_{\rm m}\cot\left[\tilde{z}/2\right]\]
with
\[  z=\rho+\mbox{i}\zeta  \quad\mbox{and}\quad
\tilde{z}=\tilde{\rho}+\mbox{i}\tilde{\zeta}.\]
The value $\rho_{\rm m}$ must be chosen such that
$\rho_1<\rho_{\rm m}<\rho_2$. 
Since in these coordinates the metric potentials are
not analytic at $\tilde{z}=0$, it is necessary to divide the corresponding compact
coordinate region of values $(\tilde{\rho},\tilde{\zeta})$ into further
subregions, each one of which is again to be mapped onto a square.

The solution is represented and determined in a completely analogous
manner as described above for the spheroidal bodies. Note that the 
initial guess for the Newton-Raphson method now comes from the
numerically known Dyson rings in Newtonian Gravity.

Again, we obtain very accurate solutions\footnote{The accuracy can be tested in
several independent ways, see \citet{An02a}.} yielding up to 12 digits for
configurations that are sufficiently far away from the 
mass-shedding limit and from the limits $\rho_1/\rho_2\to 0$,
$\rho_1/\rho_2\to 1$.
 
\section{Results}

For a particular (constant) energy density $\mu$, the
relativistic Dyson rings are characterized by two parameters, say the
redshift $Z_0$ and the radius ratio $\rho_1/\rho_2$. The region in which
these parameters may vary is depicted in Fig.~1. Here, vanishing $Z_0$
represents the Dyson rings in Newtonian Gravity. The fluid bodies with
$\rho_1/\rho_2=0$ are transition configurations from toroidal to
spheroidal topology. Starting from the Newtonian body,
these configurations reach a mass-shedding limit at $Z_0\approx 0.26$. If we now
follow the mass-shedding curve, $Z_0$ increases and reaches infinity at
$\rho_1/\rho_2=0.5613$, which corresponds to a transition to the extreme
Kerr black hole. For a rigidly rotating disk of dust, such a transition was 
conjectured by \citet{BW69,BW71} and analytically proven by
\citet{NM93,NM95}. Also for differentially rotating disks of dust a
transition of this kind has been found \citep{AnM00,Ans01}. In this limit the
coordinate extension of the gravitational source (here the
relativistic Dyson ring) shrinks until the object
coincides with the coordinate origin. The metric tensor assumes the form
of the extreme Kerr solution outside the horizon\footnote{In our coordinates,
the horizon (and the `throat' of the extreme Kerr metric) is given
by $\rho=\zeta=0$, see \citet{Mei02}.} with $W\equiv\rho$,
and the angular momentum $J$,
the gravitational mass $M$ and the angular velocity $\Omega$ approach the relation
\begin{equation}
J=M^2=(2\Omega)^{-2}.
\label{jm}
\end{equation}
This extreme Kerr black-hole limit emerges for radius ratios
$\rho_1/\rho_2>0.5613$. If we move
from here along this boundary curve towards $\rho_1/\rho_2=1$, we note
that the normalized gravitational mass
$\bar{M}=M\mu^{1/2}$, the normalized rest
mass\footnote{For the calculation of $M_0$ we
assume that the total energy density $\mu$ is equal to the rest-mass density, cf.
\citet{Bar71}.}
$\bar{M}_0=M_0\mu^{1/2}$ and the relative binding
energy
$(M_0-M)/M_0$ increase, see Fig.~2.
In the Newtonian limit ($Z_0\to 0$) the normalized angular velocity
$\bar{\Omega}=\Omega/\mu^{1/2}$ tends to zero as $\rho_1/\rho_2\to 1$. Our numerical
results support the plausible assumption that the same is valid for all values of $Z_0$.
In the black-hole limit ($Z_0\to \infty$) this would imply $\bar{M}\to\infty$ as
$\rho_1/\rho_2\to 1$, see Eq.~(\ref{jm}). Therefore we believe that there is no bound
to the gravitational mass of relativistic Dyson rings of a given mass-density as one
approaches the black-hole limit and a radius ratio of unity.

The evolution of a typical relativistic Dyson ring sequence with fixed
radius ratio $\rho_1/\rho_2=0.7$ from the Newtonian to the black-hole
limit can be seen in Figs 3 and 4. Corresponding values of various
physical quantities are given in Table 1. Finally, in Fig.~5 we provide
selected examples of configurations at the mass-shedding
and the black-hole limit. More details of our methods and results including the
discussion of realistic equations of state,
configurations with a central object (black hole or neutron star),
stability, and astrophysical relevance
will be published elsewhere.

This work was supported by the {\it Deutsche Forschungsgemeinschaft}
(DFG-project ME 1820/1-3).

\begin{table}{{\bf Table 1:} Physical quantities for the
configurations with the radius ratio $\rho_1/\rho_2=0.7$
displayed in Fig.~3. Here $\bar{\Omega}=\Omega/\mu^{1/2}$, $\bar{M}=M\mu^{1/2}$,
$\bar{M}_0=M_0\,\mu^{1/2}$, $\bar{J}=J\mu$,
and $\bar{R}_{\rm circ}=R_{\rm circ}\,\mu^{1/2}$
are normalized values of the angular velocity $\Omega$, gravitational mass $M$,
rest mass $M_0$, angular momentum $J$, and circumferential radius
$R_{\rm circ}=W{\rm e}^{-\nu}[\rho=\rho_2,\zeta=0]$.
Note that $\bar{M}$, $\bar{M}_0$, $\bar{J}$, and $\bar{R}_{\rm circ}$ tend to zero
in the Newtonian limit $Z_0\to 0$, whereas $\bar{\Omega}$, for $\rho_1/\rho_2=0.7$,
approaches the value $\bar{\Omega}_{\rm N}=0.48109$.}
   \begin{center}
    \begin{tabular}[t]{c|ccccc}
$Z_0$  & $\bar{\Omega}$  & $\bar{M}$  & $\bar{M}_0$ &  $\bar{J}$ &
$\bar{R}_{\rm circ}$\\ \hline
0.05     & 4.9108e$-$01      & 7.9661e$-$03 & 8.0842e$-$03  & 2.3168e$-$04 & 2.8483e$-$01 \\
0.50     & 5.5491e$-$01      & 1.4018e$-$01 & 1.5659e$-$01  & 2.8624e$-$02 & 7.4643e$-$01 \\
1.22     & 6.0702e$-$01      & 2.8798e$-$01 & 3.5082e$-$01  & 9.8186e$-$02 & 9.7512e$-$01 \\
1.60     & 6.2320e$-$01      & 3.3896e$-$01 & 4.2564e$-$01  & 1.3035e$-$01 & 1.0441e+00 \\
2.50     & 6.4726e$-$01      & 4.2094e$-$01 & 5.5610e$-$01  & 1.9083e$-$01 & 1.1543e+00 \\
6.00     & 6.7932e$-$01      & 5.5338e$-$01 & 7.9987e$-$01  & 3.1294e$-$01 & 1.3466e+00 \\
13.0     & 6.9211e$-$01      & 6.2871e$-$01 & 9.6211e$-$01  & 3.9769e$-$01 & 1.4734e+00 \\
$\infty$ & 6.9980e$-$01      & 7.1449e$-$01 & 1.1742e+00    & 5.1050e$-$01 & 1.6427e+00
   \end{tabular}
  \end{center}
\vspace{2cm}
\end{table}

\begin{center}
\begin{figure}
\unitlength1cm
\hspace*{-4.5cm}
\epsfig{file=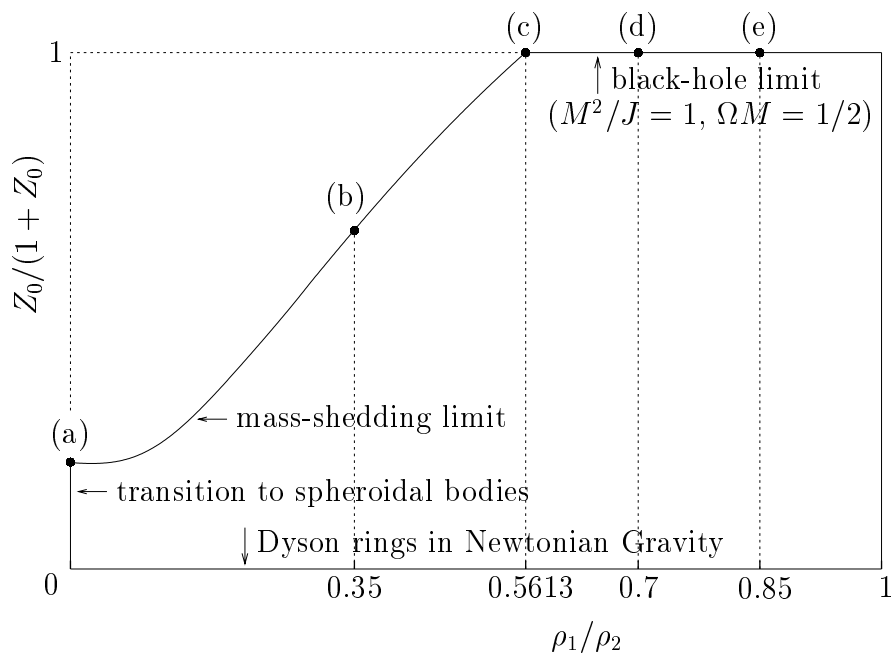,scale=1}
\caption{Parameter region of the relativistic Dyson rings in the
$\rho_1/\rho_2$ - $Z_0/(1+Z_0)$ - plane. Rings with fixed radius ratio
$\rho_1/\rho_2=0.7$ are depicted in Fig.~3, see also Fig.~4 and Table 1.
The boundary configurations indicated by $\bullet$'s can be found in Fig.~5.}
\end{figure}

\begin{figure}
\unitlength1cm
\hspace*{-2cm}
\epsfig{file=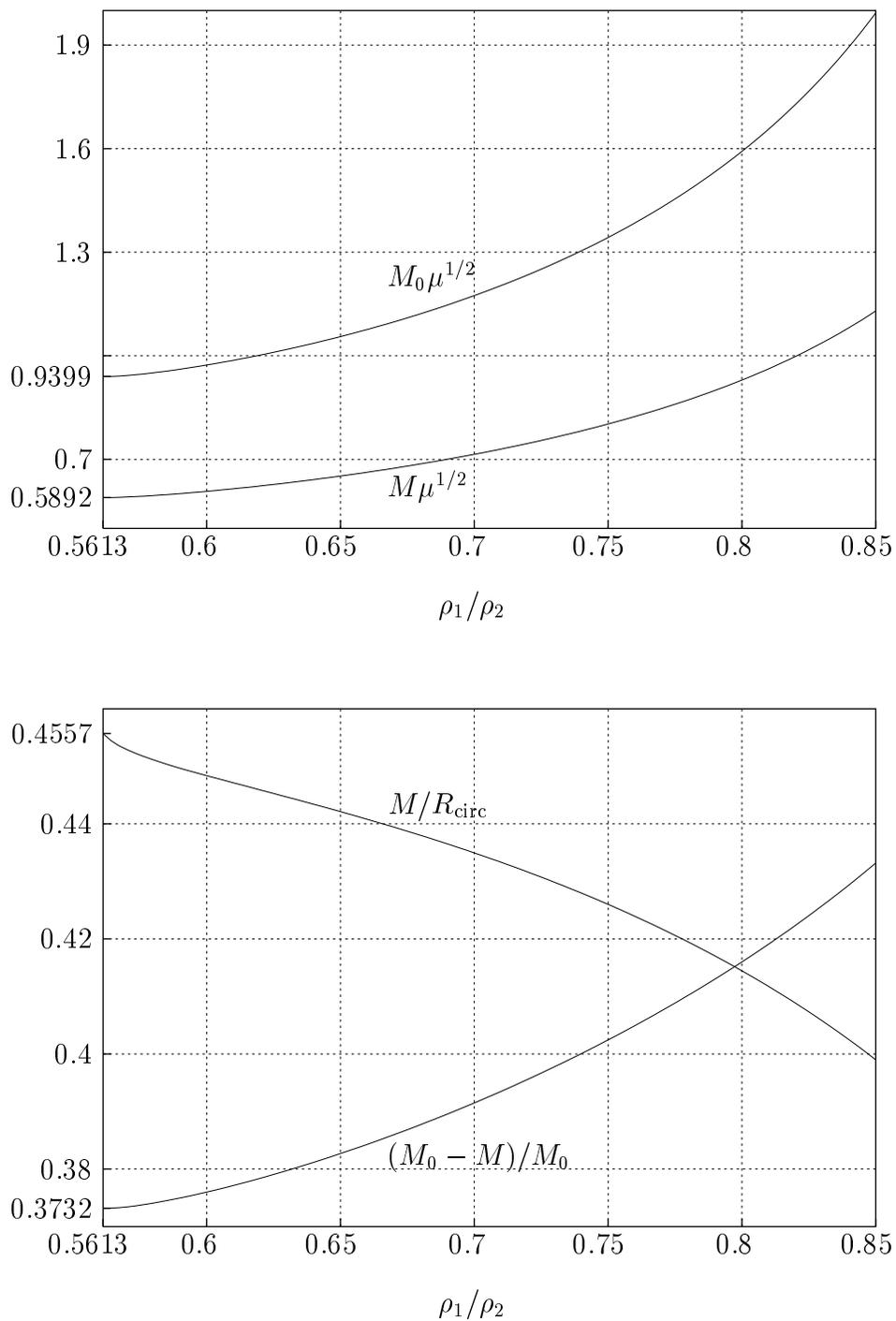,scale=1}
\caption{Normalized gravitational mass $M\mu^{1/2}$, normalized rest mass $M_0\,\mu^{1/2}$,
relative binding energy $(M_0-M)/M_0$,
and compactness parameter $M/R_{\rm circ}$ for
$0.5613\le\rho_1/\rho_2\le 0.85$ at the black-hole limit ($Z_0\to\infty$).}
\end{figure}

\begin{figure}
\unitlength1cm
\hspace*{-2cm}
\epsfig{file=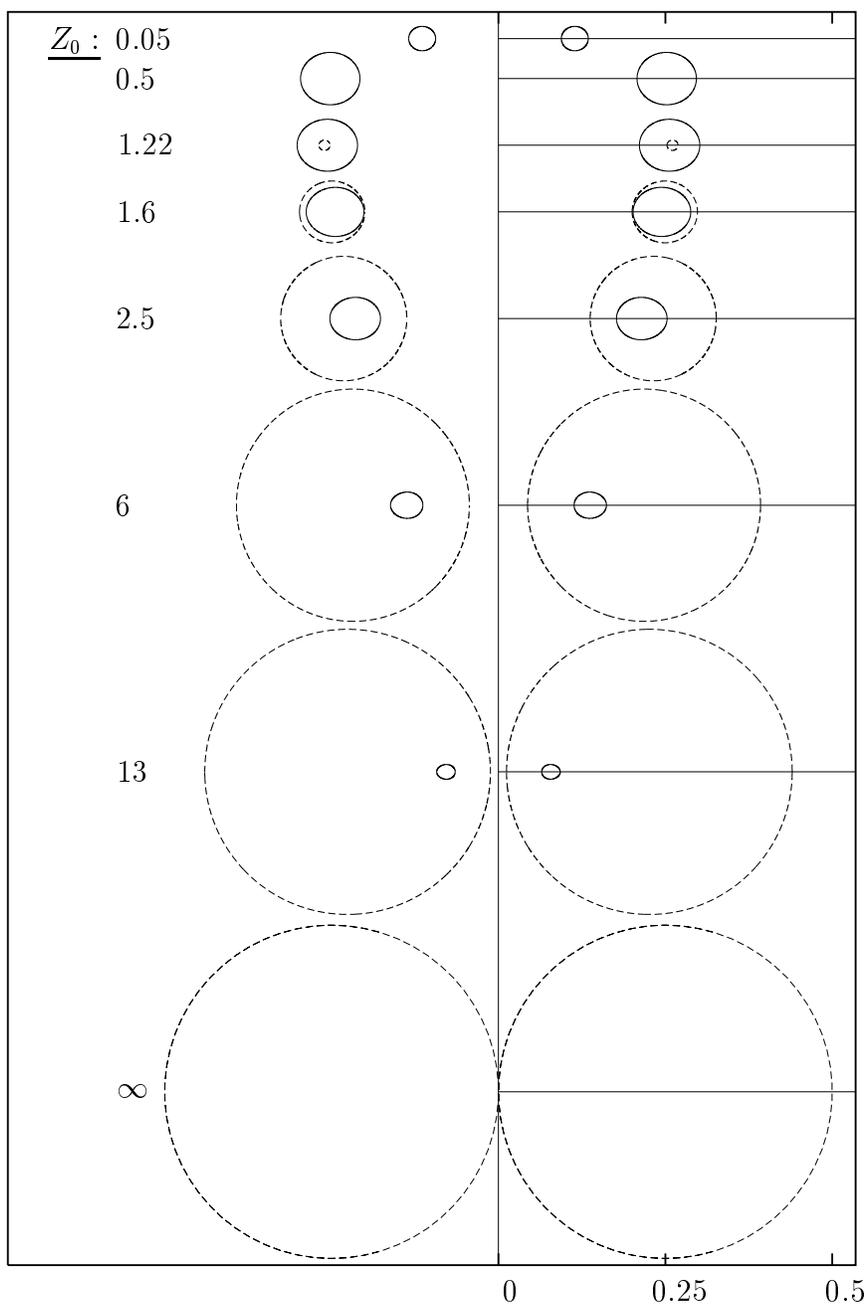,scale=1}
\caption{Cross-sections of relativistic Dyson rings with fixed coordinate-radius
ratio $\rho_1/\rho_2=0.7$ and varying redshift $Z_0$. The normalized $\zeta$-coordinate
$\Omega\zeta$ is plotted against the normalized $\rho$-coordinate $\Omega\rho$ (with the
axes scaled identically).
In the Newtonian limit ($Z_0\to 0$)
as well as in the black-hole limit ($Z_0\to\infty$) the ring shrinks down to the
normalized coordinate origin. Dashed lines represent the boundary of the toroidal ergo-region.
In the limit $Z_0\to\infty$ the ergo-region is that of the extreme Kerr black hole,
cf. \citet{Mei02}.}
\end{figure}

\begin{figure}
\unitlength1cm
\hspace*{-2.5cm}
\epsfig{file=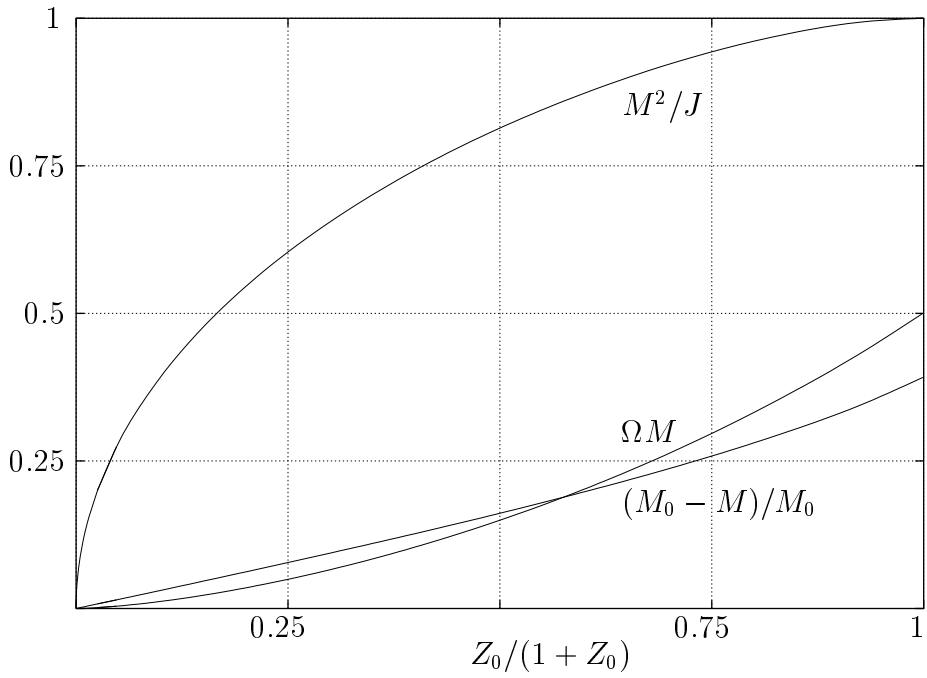,scale=1}
\caption{The dimensionless quantities $M^2/J$, $\Omega M$ and $(M_0-M)/M_0$ for the relativistic
Dyson ring sequence with $\rho_1/\rho_2=0.7$ from the Newtonian limit ($Z_0\to 0$) to
the black-hole limit ($Z_0\to\infty$).}
\end{figure}

\begin{figure}
\unitlength1cm
\hspace*{-2cm}
\epsfig{file=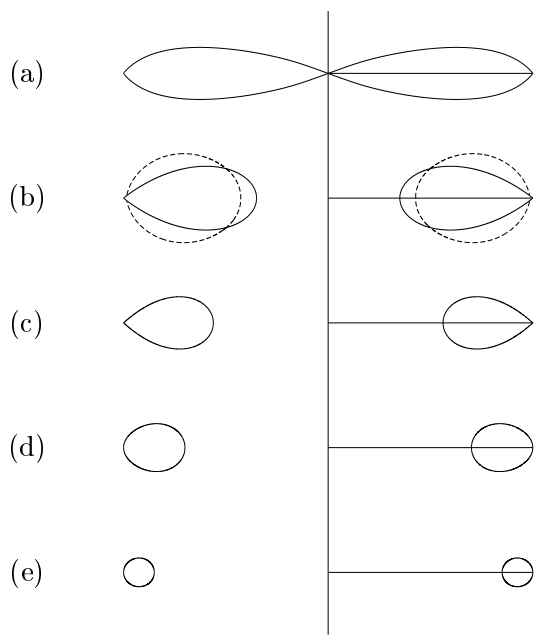,scale=1}
\caption{Cross-sections of relativistic Dyson rings at the mass-shedding limit (a) -- (c) and
at the black-hole limit (c) -- (e), cf. Fig.~1. Here the normalized $\zeta$-coordinate
$\zeta/\rho_2$ is plotted against the normalized $\rho$-coordinate $\rho/\rho_2$ (with the
axes scaled identically). The dashed cross-section in (b) again represents the ergo-region.
Note that $\Omega\rho_2\to 0$ in the black-hole limit, see Fig.~3.
Hence, in contrast to (b), the ergo-region would become infinitely
large with the scaling of this figure. It is interesting to observe a more
and more circular shape of the ring's cross-section as one approaches $\rho_1/\rho_2=1$,
a property known from the Dyson rings in Newtonian Gravity.}
\end{figure}
\end{center}

\end{document}